\documentclass[preprint]{aastex}
\usepackage{graphicx}
\usepackage{amssymb}

\shorttitle{XMM-Newton X-Ray Observation of PSR J1734--3333}
\shortauthors{Olausen et al.}

\begin{document}

\title{\textit{XMM-Newton} X-ray Observation of the High-Magnetic-Field
Radio Pulsar PSR J1734--3333}

\author{S. A. Olausen\altaffilmark{1}, V. M. Kaspi\altaffilmark{1}, A.
G. Lyne\altaffilmark{2}, and M. Kramer\altaffilmark{2,3}}

\altaffiltext{1}{Department of Physics, Rutherford Physics Building, McGill University,
3600 University Street, Montreal, Quebec, H3A 2T8, Canada}

\altaffiltext{2}{Jodrell Bank Centre for Astrophysics, School of Physics and Astronomy,
University of Manchester, Manchester, M13 9PL, UK}

\altaffiltext{3}{MPI f\"ur Radioastronomie, Auf dem H\"ugel 69, 53121 Bonn,
Germany}
\begin{abstract}
Using observations made with the \textit{XMM-Newton} Observatory,
we report the probable X-ray detection of the high-magnetic-field
radio pulsar PSR J1734--3333. This pulsar has an inferred surface
dipole magnetic field of $B=5.2\times10^{13}\,\mathrm{G}$, just below
that of one anomalous X-ray pulsar (AXP). We find that the pulsar
has an absorbed 0.5--2.0\,keV flux of $\left(5\textrm{--}15\right)\times10^{-15}\,\mathrm{erg\, s^{-1}\, cm^{-2}}$
and that its X-ray luminosity $L_{\mathrm{X}}$ is well below its
spin down luminosity $\dot{E}$, with $L_{\mathrm{X}}<0.1\dot{E}$.
No pulsations were detected in these data although our derived upper
limit is unconstraining. Like most of the other high-$B$ pulsars,
PSR J1734--3333 is X-ray faint with no sign of magnetar activity.
We tabulate the properties of this and all other known high-$B$ radio
pulsars with measured thermal X-ray luminosities or luminosity upper
limits, and speculate on a possible correlation between
$L_{\mathrm{X}}$ and $B$.
\end{abstract}


\section{Introduction}

Soft gamma repeaters (SGRs) and anomalous X-ray pulsars (AXPs) are
now well accepted as being different though similar manifestations
of {}``magnetars'' --- isolated neutron stars whose radiation is
powered by their magnetic field (see \citealt{wt06} or \citealt{m08}
for a review). These objects are characterised by long spin periods
(2--12\,s), spin-down rates that, assuming conventional magnetic
dipole braking, imply surface dipolar magnetic fields in the range
$\sim$$10^{14}\textrm{--}10^{15}\,\mathrm{G}$, X-ray luminosities
of $10^{33}\textrm{--}10^{34}\,\mathrm{erg\, s^{-1}}$ that are orders
of magnitude greater than their spin-down luminosities, and great
X-ray variability, ranging from short SGR-like bursts to major, slow-rise
and long-lived X-ray flares (see \citealt{k07} for a review). The
bulk of SGR and AXP properties are well explained by the magnetar
model \citep{td95,td96,tlk02}.

However, there remain some outstanding challenges to the magnetar
picture. Arguably the most important is the connection between magnetars
and high-magnetic field radio pulsars. Some models predict that given a
large enough magnetic field $\bigl($presumably close to the quantum
critical field, $B_{\mathrm{QED}}=4.4\times10^{13}\,\mathrm{G}\bigr)$,
conventional radio emission should be suppressed \citep{bh98}. Yet there
are now known over a half dozen otherwise ordinary radio pulsars having
spin-down inferred magnetic field $B>4\times10^{13}\,\mathrm{G}$ $\bigl($where
$B=3.2\times10^{19}\bigl(P\dot{P}\bigr)^{1/2}\,\mathrm{G}$ for a
pulsar with period $P$ and spin-down rate $\dot{P}\bigr)$, with
none showing evidence for conventional magnetar-like emission. Some,
such as PSRs J1847--0130 and J1718--3718, have $B$ greater than those
measured for \textit{bona fide} magnetars, yet no anomalous X-ray
emission \citep{msk+03,km05}. Uncertainties in inferred $B$ from
spin-down can be substantial \citep[e.g.,][]{hck99,s06}, but still,
in the magnetar picture, some evidence for anomalous X-ray emission
is reasonably expected in some high-$B$ radio pulsars. This is important
overall for unifying the surprisingly varied manifestations of neutron
stars into a physical framework (see \citealt{k10} for a review).

Recently, the idea that high-$B$ pulsars might exhibit anomalous
X-ray emission was confirmed by the discovery of SGR-like X-ray bursts
and a long-lived X-ray flux enhancement from what was previously thought
to be a purely rotation-powered pulsar. \citet{ggg+08} found that
the high-$B$ pulsar PSR J1846--0258 at the center of the supernova remnant
Kes 75 emitted several SGR-like bursts in 2006, contemporaneous with a
flux enhancement and a rotational glitch \citep[see also][]{ks08,nsgh08,lkg10}.
This is the first pulsar known to have quiescent X-ray luminosity that
could be rotation-powered, and indeed has many properties of rotation-powered
pulsars, while showing obvious magnetar-like behavior. Its small
characteristic age of 884\,yr lends further credence to the idea --- PSR
J1846--0258 could be a very young magnetar, and one of the {}``missing links''
in the hypothesized high-$B$ pulsar/magnetar evolutionary chain.

PSR J1734--3333 is a radio pulsar with $P=1.169\,\mathrm{s}$ and
$\dot{P}=2.3\times10^{-12}$, which imply a spin-down luminosity of
$\dot{E}=4\pi^{2}I\dot{P}/P^{3}=5.6\times10^{34}\,\mathrm{erg\, s^{-1}}$
and a characteristic age of $\tau=P/2\dot{P}=8.1\,\mathrm{kyr}$.
Its inferred surface dipolar magnetic field is $B=5.2\times10^{13}\,\mathrm{G}$,
which is among the highest of all known radio pulsars and just below
those of \textit{bona fide} magnetars such as 1E 2259+586 $\left(B=5.9\times10^{13}\,\mathrm{G}\right)$.
From 12\,yr of phase-coherent radio timing observations at Jodrell
Bank, this pulsar is known to have a stable braking index measured of
$n=1.0\pm0.3$ \citep{elk+11}. At face value, this could imply that
this high-$B$ pulsar's magnetic field is growing, i.e.,\ its trajectory
on a conventional $P/\dot{P}$ diagram is up and to the right, toward
the region occupied by the magnetars. This makes PSR J1734--3333 a
good candidate for exhibiting magnetar-like anomalous X-ray emission.
For this reason, we obtained \textit{XMM-Newton} observations of this
source which we report on here.

\section{Observations and Results}

One 10-ks observation of PSR J1734--3333 was carried out on 2009 March
9-10 using the \textit{XMM-Newton} Observatory \citep{jla+01}. The
two EPIC MOS cameras \citep{taa+01} were operating in full-window
mode with a time resolution of 2.7\,s, and the EPIC pn camera \citep{sbd+01}
was in large-window mode providing a time resolution of 48\,ms. For
all three cameras the medium filter was in use. The total exposure
time in this observation was 10.6\,ks with each MOS camera and 8.7\,ks
with the pn.

The data were analyzed with the \emph{XMM} Science Analysis System
(SAS) version 8.0.0%
\footnote{See \url{http://xmm.esac.esa.int/sas/8.0.0/}%
} with calibrations updated 2009 April 13. To identify times contaminated
by strong background flaring that is sometimes present in \textit{XMM}
data, we extracted and examined light curves of photons above 10\,keV
over the entire field of view for the pn and MOS cameras. No flaring
or significant background changes were found, so the entire exposure
could be used in our analysis.

\subsection{Imaging}

In order to find a possible X-ray counterpart of PSR J1734--3333,
we performed a blind search for point sources using the SAS tool \texttt{edetect\_chain}.

A faint X-ray source was detected near the radio position of the pulsar
by \texttt{edetect\_chain} in all three cameras. In the pn camera
it had $49\pm10$ counts in the 0.5--3.0\,keV energy range and a
likelihood ratio of $L_{2}=-\ln P\simeq29$ (where $P$ is the probability
that a random Poissonian fluctuation could produce the observed source
counts). The source was also detected, in the same 0.5--3.0\,keV
energy band, in the MOS 1 camera with $21\pm6$ counts and $L_{2}\simeq12$
and in the MOS 2 camera with $31\pm8$ counts and $L_{2}\simeq17$.
Outside this energy range there were almost no source counts detected.

Figure~\ref{fig:Image} shows the X-ray emission near the radio position
of the pulsar, made by combining the pn, MOS 1, and MOS 2 images into
a mosaic and smoothing with a Gaussian kernel of radius $\sigma=3\arcsec$.
The best-fit position of the X-ray source as reported by \texttt{edetect\_chain}
is (J2000) $\mathrm{R.A.}=17^{\mathrm{h}}34^{\mathrm{m}}27\fs19\pm0\fs24$,
$\mathrm{DECL.}=-33\degr33\arcmin22\farcs0\pm3\farcs0$, where these
uncertainties consist of both the $1\arcsec$ statistical error and
\textit{XMM}'s absolute pointing uncertainty%
\footnote{See \url{http://xmm2.esac.esa.int/docs/documents/CAL-TN-0018.pdf}%
} of $2\arcsec$. In principle it is possible to reduce the pointing
uncertainty by matching at least two bright X-ray sources in the field
to known optical counterparts; unfortunately, the field of view of
this observation contained only one such source.

The most up to date radio timing position for PSR J1734--3333 is $\mathrm{R.A.}=17^{\mathrm{h}}34^{\mathrm{m}}26\fs9\pm0\fs2$,
$\mathrm{DECL.}=-33\degr33\arcmin20\arcsec\pm10\arcsec$ \citep{elk+11}.
The offset between the radio timing position and the centroid of the
X-ray position in R.A. is $0\fs29$ or $\sim$$0.9\,\sigma$, and
is $2\farcs0$ or $\sim$$0.2\,\sigma$ in declination.

\subsection{Spectroscopy}

To obtain the X-ray spectrum, we extracted a total of 300 photon events
from the pn data using a circular region of $30\arcsec$ radius centered
on the source. The background spectrum was taken from an elliptical
annulus with semi-major and semi-minor axes of $75\arcsec$ and $45\arcsec$
rotated $90^{\circ}$ from north, surrounding but excluding the $30\arcsec$
source region. This background region contained 643 events, implying there
to be $\sim$234 background photons in the source region. A response and
ancillary response file were generated using the SAS tasks \texttt{rmfgen}
and \texttt{arfgen}. The spectrum was grouped to have a minimum of 20
counts per bin with the \texttt{ftool} \texttt{grppha} and was fed into
XSPEC 12.5.0ac for spectral fitting. We did not extract spectra from the
MOS images due to the low number of source counts in those cameras.

\begin{figure}
\includegraphics[angle=270,scale=0.5]{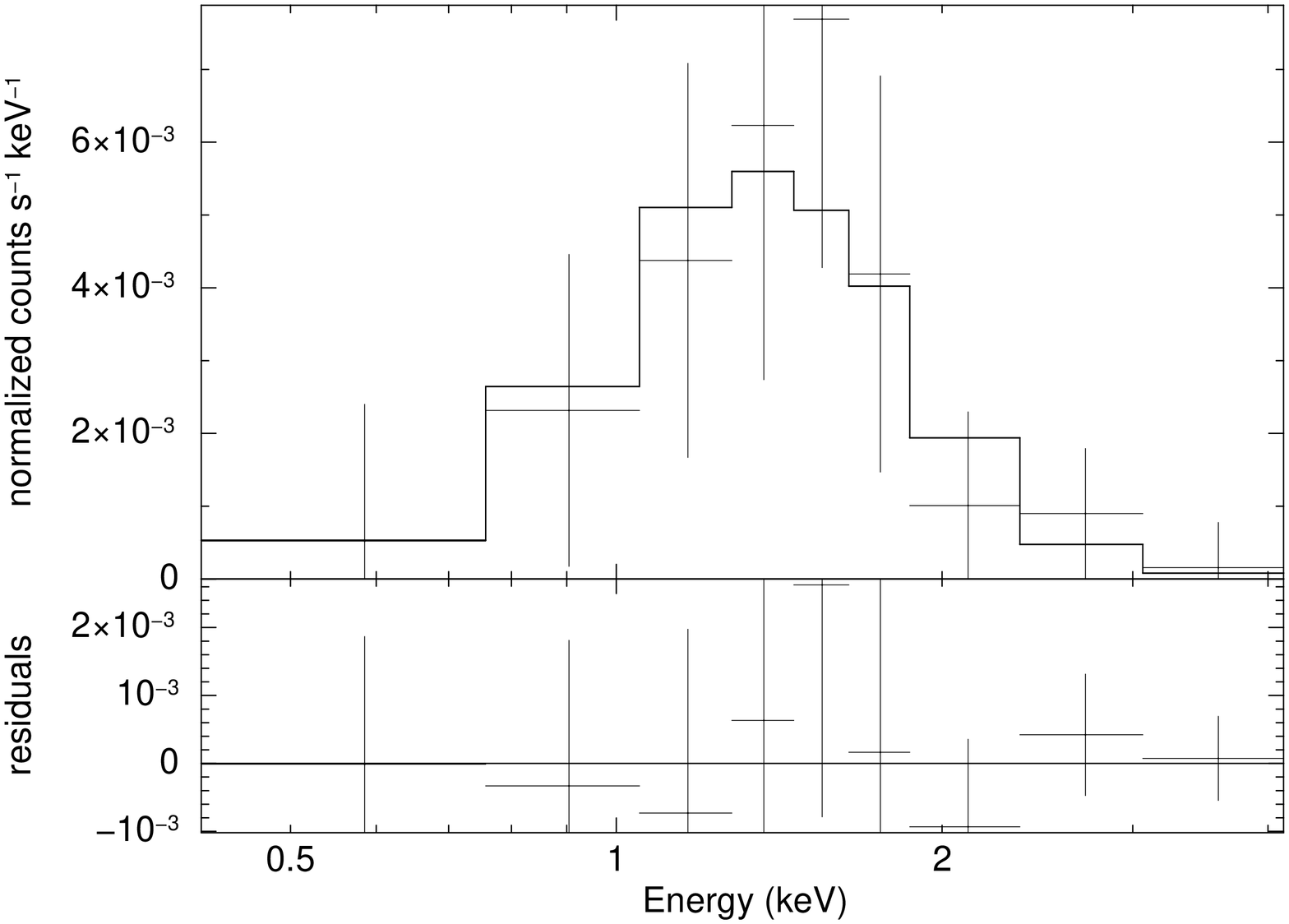}

\includegraphics[angle=270,scale=0.5]{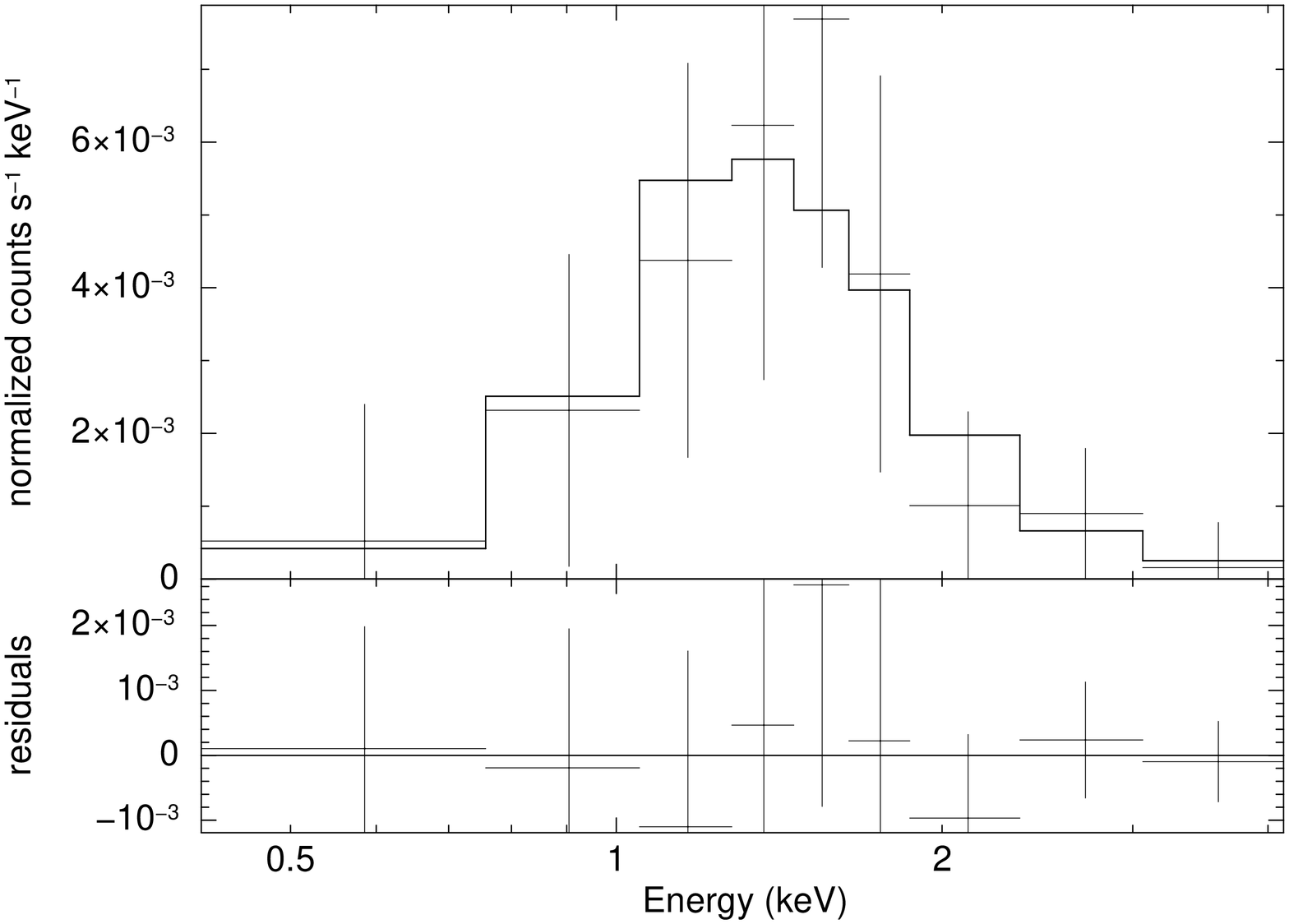}

\caption{\textit{XMM} pn spectrum of the possible X-ray counterpart to PSR
J1734--3333. The spectrum is binned to contain a minimum of 20 counts
per bin. Top: best-fit blackbody model for fixed $N_{\mathrm{H}}=0.6\times10^{22}\,\mathrm{cm^{-2}}$.
Bottom: best-fit power-law model for fixed $N_{\mathrm{H}}=1.2\times10^{22}\,\mathrm{cm^{-2}}$.\label{fig:Spectrum}}

\end{figure}

The X-ray spectrum was fit to absorbed power-law and absorbed blackbody
models. Because of the small number of source counts, it was impossible
to constrain the column density $N_{\mathrm{H}}$. The Leiden/Argentine/Bonn
and Dickey \& Lockman Surveys of Galactic H~\textsc{i}%
\footnote{\url{http://cxc.harvard.edu/toolkit/colden.jsp}%
} give the total column density along the line of sight to the pulsar
as $N_{\mathrm{H}}=\left(1.1\textrm{--}1.4\right)\times10^{22}\,\mathrm{cm}^{-2}$,
and from its dispersion measure of $578\,\mathrm{pc\, cm}^{-3}$ the
distance to the pulsar is estimated to be 6.1\,kpc with a $\sim$25\%
uncertainty \citep{cl01}. We therefore assumed an upper limit on
$N_{\mathrm{H}}$ of $1.2\times10^{22}\,\mathrm{cm}^{-2}$, halved
it to get a more reasonable estimate of $0.6\times10^{22}\,\mathrm{cm}^{-2}$
since, based on its distance, the pulsar is less than halfway through
the Galaxy, and found best-fit models after fixing $N_{\mathrm{H}}$
to both values. The parameters of these best-fit models are shown
in Table~\ref{tab:Spectral}, and two of them are plotted in Figure~\ref{fig:Spectrum}.
In general, a lower value of $N_{\mathrm{H}}$ corresponds to a power-law
model with smaller photon index and a blackbody model with a higher
$kT$.

\subsection{Timing Analysis}

Radio pulsar timing data were obtained using the Jodrell Bank
76-m Lovell telescope operating at a central observing frequency of
around 1400\,MHz, roughly contemporaneously with the X-ray observations
(see \citealt{elk+11} for details). The TEMPO timing package%
\footnote{See \url{http://www.atnf.csiro.au/research/pulsar/tempo/}%
} was used to correct the times of arrival to the barycenter of the
solar system assuming the position determined by \citet{elk+11} and
to fit a rotational model to these arrival times.

Based on this ephemeris for PSR J1734--3333, during the \textit{XMM}
observation the pulsar had an expected barycentric frequency of $854.9300124\pm0.0000004\,\mathrm{mHz}$.
Since the MOS cameras were operated in full-window mode with 2.7-s
time resolution, their data cannot be used for our timing analysis.
A search for X-ray pulsations was done using the pn data after barycentering
them using the SAS tool \texttt{barycen}. To do so, we extracted counts
in the 0.5--3.0\,keV energy range from a region of $20\arcsec$ radius
centered on the source. Although we used a smaller source region
here than for the spectral analysis in order to maximize the
signal-to-noise ratio, an identical background region was used for both
the timing and spectral work. This background region contained 286
photon events in the same 0.5--3.0\,keV energy band, implying that, of
the 84 total events in the source region, $\sim$46 were background photons.
The source region event list was folded at the radio frequency using 16
phase bins, and the folded light curve was fit to a constant line. The
best-fit $\chi^{2}$ was 12.2 for 15 degrees of freedom, corresponding to
a 67\% probability that the folded curve could be produced from a data
set containing no signal. Thus, no significant pulsations were detected.
Searches were also conducted in the 0.5--10\,keV and 1--2\,keV energy
bands with similar null results.

To find an upper limit for the pulsed fraction, we simulated event
lists with the same number of total counts as found in the source
region. The simulated signal had a sinusoidal profile with a random
phase and had a user specified area pulsed fraction, where the area
pulsed fraction is defined as the ratio of the pulsed part of the
profile to the entire profile. However, given the expected number
of background counts, even a signal with 100\% area pulsed fraction
could only be detected with $>$$3\sigma$ significance 5\% of the
time. Thus, we conclude that there are too few counts in our data
set to set a meaningful upper limit on the pulsed fraction.

\section{Discussion}

Although the location of the detected X-ray source is consistent with the
radio timing position of PSR J1734--3333, given that no X-ray pulsations
were detected from it at the radio period to provide unambiguous proof of
association, it is reasonable to question whether the X-ray source really
is associated with the radio pulsar. We can estimate the probability of a
chance superposition from the $\log N\textrm{--}\log S$ curves for \textit{XMM-Newton}
Galactic-plane sources in the 0.5--2.0\,keV band \citep{m06}. Given
the lowest reasonable value of absorbed flux in that band for the
source, $\sim$$5\times10^{-15}\,\mathrm{erg\, s^{-1}\, cm^{-2}}$,
the $\log N\textrm{--}\log S$ curves predict $\sim$180 sources per
square degree at this flux or higher. The probability of a random
X-ray source lying within the radio error ellipse is then only 0.1\%--0.3\%.

\begin{figure}
\includegraphics{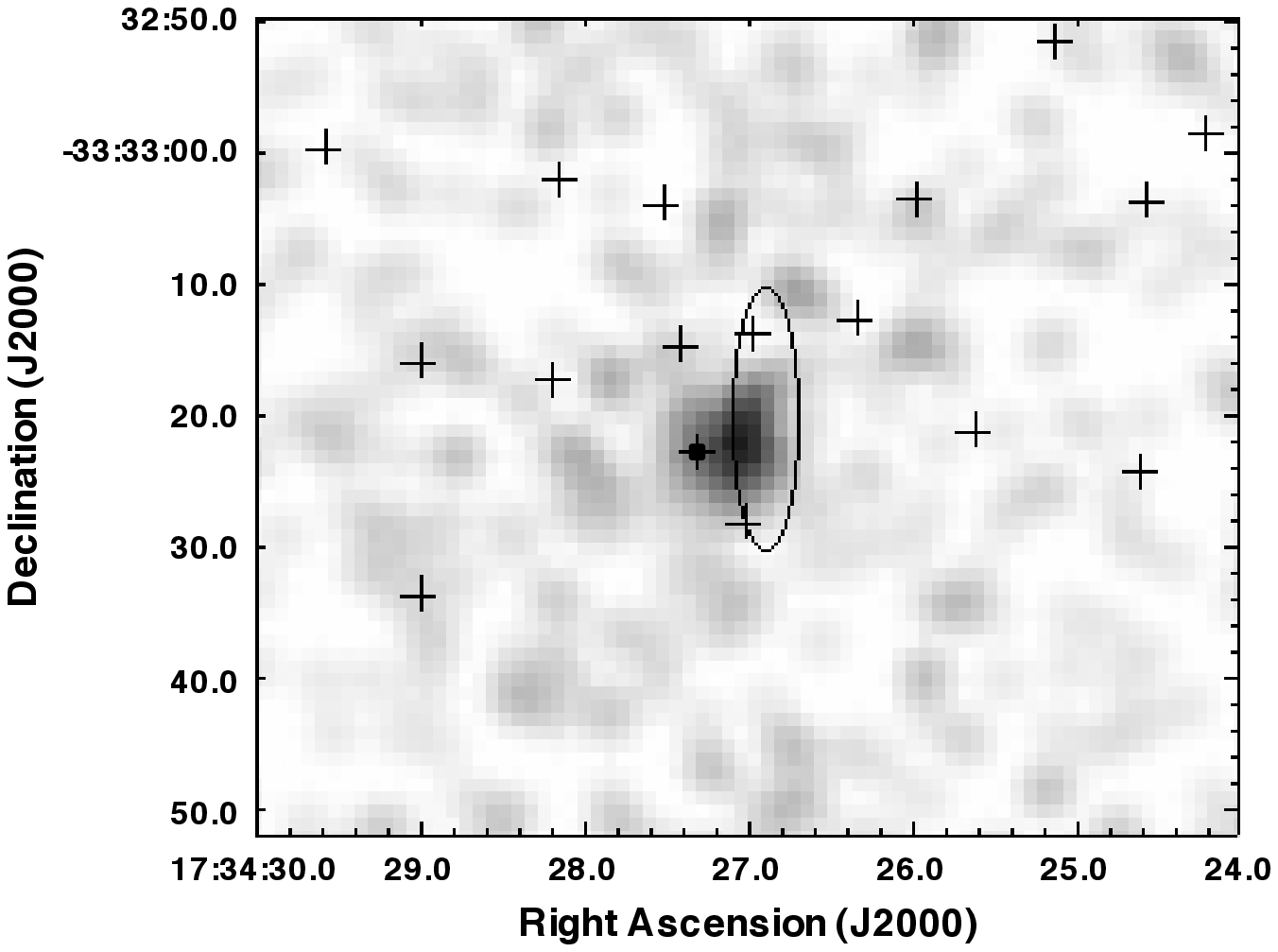}

\caption{\textit{XMM} image of the PSR J1734--3333 field in the $0.5-3.0\,\mathrm{keV}$
band, smoothed by a Gaussian kernel with $\sigma=3\arcsec$. The radio
timing position is shown by the ellipse, and the crosses denote the
positions of optical sources from the USNO-B1.0 catalog. The optical
source closest to the X-ray source (NOMAD Catalogue ID 0564-0621454)
is represented by the cross marked with a box.\label{fig:Image}}

\end{figure}

On the other hand, looking at an optical image of the field of the
\textit{XMM} observation, there are several sources near the X-ray
source. In fact, one of these optical sources (NOMAD Catalog ID 0564-0621454)
lies within the X-ray error circle at coordinates $\mathrm{R.A.}=17^{\mathrm{h}}34^{\mathrm{m}}27\fs319\pm0\fs008$,
$\mathrm{DECL.}=-33^{\circ}33^{\prime}22\farcs61\pm0\farcs05$ (See
Figure~\ref{fig:Image}). Therefore the possibility must be considered that
the X-ray source is associated with this optical source in addition to
or instead of the radio pulsar.

Assuming that the X-ray source is associated with 0564-0621454, its
X-ray to optical flux ratio can be estimated using the following formula
from \citet{ggv+04}:\[
\log\left(f_{\mathrm{X}}/f_{\mathrm{opt}}\right)=\log f\left(\textrm{0.5--8\,\ keV}\right)+0.4B+4.89.\]
Here $f\left(\textrm{0.5--8\,\ keV}\right)$ is the unabsorbed X-ray
flux in that energy band and $B=18.79$ is the optical magnitude of
0564-0621454 taken from the USNO-B1.0 catalog. Although the unabsorbed
X-ray flux is even less constrained than the absorbed flux used above,
we can calculate a rough upper limit of $\log\left(f_{\mathrm{X}}/f_{\mathrm{opt}}\right)<1$.
PSR J1734--3333 is an isolated neutron star, however, which typically
has $\log\left(f_{\mathrm{X}}/f_{\mathrm{opt}}\right)\sim5$ \citep{ttzc00,krh06}.
Therefore, we conclude that if our X-ray source is associated with
0564-0621454, it cannot also be associated with PSR J1734--3333, and
vice versa.

Finally, just as the probability of a chance superposition of an X-ray
source with the pulsar's radio position was estimated above, we can
calculate the probability of an optical source coinciding with the X-ray
position. A query of the USNO-B1.0 catalog returns almost 5500 sources
within a $10\arcmin$ radius of the centre of the \textit{XMM} field.
The probability of one of these sources lying within the $3\arcsec$ X-ray
error circle by chance is $\sim$12\%, much higher than the probability
of a chance X-ray/radio alignment. Even accounting for the fact that
0564-0621454 is only $2\arcsec$ away from the X-ray position, the
probability of a chance superposition is still 5\%--6\%. We conclude that
it is highly likely that the X-ray source is associated with PSR
J1734--3333, and much less probable that it is associated with the
optical source 0564-0621454.

With so few source counts to work with, the X-ray source's spectrum allows
for such a wide range of possible photon indices and blackbody temperatures
that it is impossible to eliminate a thermal or non-thermal source
for the X-ray emission. Among the four models shown in Table~\ref{tab:Spectral},
the absorbed 0.5--2.0 keV flux is relatively well constrained, to
within a factor of 2--3. However, since assuming a higher value of
$N_{\mathrm{H}}$ yields a steeper power-law index or a lower $kT$,
the unabsorbed 0.5--2.0 keV flux is much less constrained and varies
by nearly 2 orders of magnitude within and between the models. Assuming
a distance to the pulsar of 6.1\,kpc as estimated from the dispersion
measure, if the X-ray source is the pulsar, the 0.5--10.0\,keV X-ray
luminosity of PSR J1734--3333 is in the range $\left(1\textrm{--}34\right)\times10^{32}\,\mathrm{erg\, s^{-1}}$.
However, $L_{\mathrm{X}}<\dot{E}$ for all these models, even allowing
for a 25\% uncertainty in the estimated distance, with $L_{\mathrm{X}}/\dot{E}\sim0.1\%\textrm{--}10\%$.
Furthermore, the high end of the luminosity range corresponds to the
high $N_{\mathrm{H}}$ models, suggesting that the pulsar's true X-ray
luminosity more likely lies in the lower end of the range, if the
association holds. If it does not hold, the pulsar is even less luminous.

The best-fit blackbody temperature, $kT=0.25_{-0.08}^{+0.13}\,\mathrm{keV}$,
found for this pulsar is also unusually high for its age (0.07--0.11\,keV;
\citealt{pgw06}). \citet{plmg07} reported a possible correlation between
blackbody temperature and magnetic field, which, given the high $B$-field
of PSR J1734--3333, could explain the discrepancy. Nevertheless, the
uncertainties on $kT$ are very large. Additionally we have not taken
into account the effect of the neutron star atmosphere, and spectral
models that do include the atmosphere result in a lower effective
temperature. As such, without better statistics, we cannot eliminate the
possibility that the pulsar's blackbody temperature is within the
expected limits. 

In Table~\ref{tab:HighB} we list all high-magnetic field radio pulsars
with inferred magnetic field $B>B_{\mathrm{QED}}$, as well as all other radio
pulsars with $B>1.5\times10^{13}\,\mathrm{G}$, that have measured
thermal X-ray luminosities or luminosity upper limits. Of these stars
only one, PSR J1846--0258, has exhibited clear magnetar-like behavior
\citep{ggg+08}. Other than that, the high-$B$ pulsars show X-ray spectral
properties that are much different from those of active magnetars, being
much softer and fainter X-ray sources. They do, however, show some
similarities with transient magnetars, which in quiescence are also
softer, fainter X-ray sources. One transient magnetar in particular,
XTE J1810--197, displays spectral properties in quiescence that are
consistent with those of the detected high-$B$ pulsars $\left(kT\approx0.18\,\mathrm{keV}\textrm{ and 0.5--10\,\ keV }L_{\mathrm{X}}\sim7\times10^{32}\,\mathrm{erg\, s^{-1}}\right)$
\citep{ghbb04}.

\begin{figure}
\includegraphics[scale=0.66]{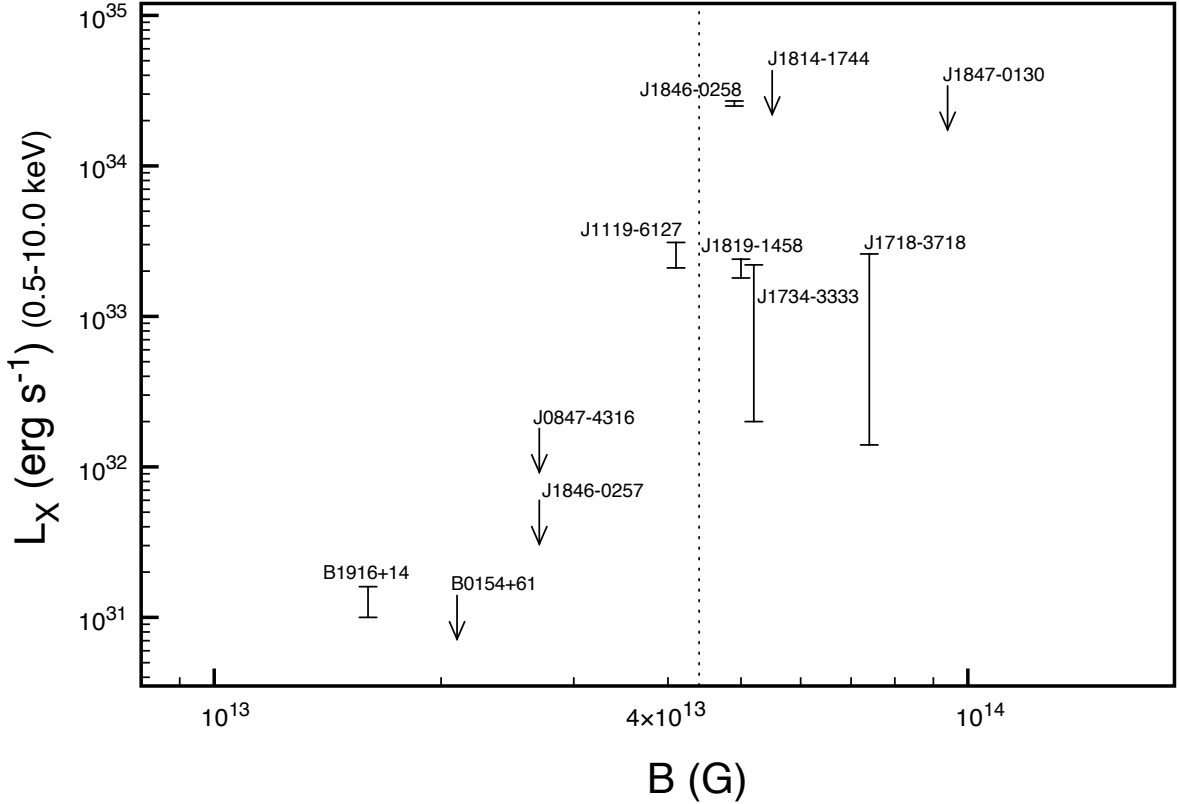}

\caption{X-ray luminosity (0.5--10\,keV) vs.\ inferred surface dipolar
magnetic field strength for X-ray observed high-$B$ radio pulsars.
Luminosity values from the original references were extrapolated to this
energy band based on the fit or assumed spectral model. Error bars are
at the $1\sigma$ confidence level and do not include uncertainties in the
distance measurements. Note that the value plotted for PSR J1846--0258
is its quiescent X-ray luminosity. The dotted line represents the quantum
critical field, $B_{\mathrm{QED}}=4.4\times10^{13}\,\mathrm{G}$. See
Table~\ref{tab:HighB} for references.\label{fig:LXvB}}

\end{figure}

Additionally, we took the reported X-ray luminosities for the pulsars
in Table~\ref{tab:HighB}, converted them to the 0.5--10\,keV
energy band, and plotted this value versus inferred magnetic field
in Figure~\ref{fig:LXvB}. The resulting plot shows possible evidence
of a correlation between the two quantities, with higher $B$-field
stars having higher 0.5--10\,keV $L_{\mathrm{X}}$; certainly the
pulsars here with $B\gtrsim4\times10^{13}\,\mathrm{G}$ are brighter
in X-rays than those with $B\lesssim4\times10^{13}\,\mathrm{G}$.
There are important caveats that should be noted, however. In particular,
the four youngest pulsars in Table~\ref{tab:HighB} also comprise four
out of the five brightest X-ray detected pulsars in the Table,
suggesting that the perceived correlation may be due entirely or in
part to age instead of $B$. On the other hand, one of the brightest of
these pulsars, PSR J1819--1458, has a considerably higher characteristic
age; it is higher even than that of the much dimmer PSR B1916+14.
Additionally, of the four pulsars with the highest $B$ fields, two have
only X-ray luminosity upper limits and the other two have only weak
detections with poorly constrained luminosity measurements. Future
observations may show some or all of these pulsars to be substantially
more X-ray dim than PSR J1119--6127 or PSR J1819--1458, which would
considerably weaken the possible correlation.

Although the spectrum of PSR J1734--3333 is poorly constrained, it
is evident that it has much more in common with the other high-$B$
rotation-powered pulsars than with the magnetars. Most magnetars, even
in quiescence, have X-ray luminosities exceeding their spin-down
luminosities by orders of magnitude; for this pulsar, as with the other
high-$B$ radio pulsars, the opposite is true. In particular, its
2--10\,keV X-ray luminosity is well below that of any known magnetar
\citep[see][Table 14.1]{wt06}. Therefore, while PSR J1734--3333 could be
argued to be evolving towards the rotational properties of a magnetar
\citep{elk+11}, it currently retains essentially the X-ray radiative
properties of a radio pulsar. What is less clear from these data,
however, is whether its 0.5--10\,keV $L_{\mathrm{X}}/\dot{E}$ is similar
to or greater than that of PSR J1846--0258 in quiescence ($\sim$1\%--10\%),
or whether it is considerably smaller as with most other high-$B$ pulsars.
Unfortunately, due to the poor statistics it is difficult to compare the
spectrum of PSR J1734--3333 with those of other high-$B$ pulsars and
magnetars. Longer X-ray observations are needed to better constrain the
luminosity, obtain a good spectrum, and search for X-ray pulsations.

\acknowledgements{We thank R. Rutledge and M. Livingstone for useful
comments. V.M.K. receives support from NSERC, FQRNT, CIFAR, and holds
a Canada Research Chair and the Lorne Trottier Chair in Astrophysics
and Cosmology.}

\clearpage{}

\begin{deluxetable}{ccccc}
\tablewidth{0pt}
\tablecolumns{5}
\tablecaption{Spectral Parameters for PSR J1734--3333\label{tab:Spectral}}
\tablehead{\colhead{Parameter} & \multicolumn{2}{c}{Power Law} & \multicolumn{2}{c}{Blackbody}}

\startdata
$N_{H}$ $\left(10^{22}\,\mathrm{cm}^{-2}\right)$ & 1.2 & 0.6 & 1.2 & 0.6\\
$\Gamma$\tablenotemark{a} & $4.2_{-1.4}^{+1.8}$ & $2.6_{-1.0}^{+1.3}$ & \nodata & \nodata\\
$kT$ (keV)\tablenotemark{a} & \nodata & \nodata & $0.25_{-0.08}^{+0.13}$ & $0.34_{-0.11}^{+0.19}$\\
$\chi^{2}$ (dof) & 1.44(7) & 2.55(7) & 1.35(7) & 1.49(7)\\
$f_{\mathrm{abs}}$ $\left(10^{-15}\,\mathrm{erg\, s^{-1}\, cm^{-2}}\right)$\tablenotemark{b} & 6.4--16 & 5.8--15 & 5.7--15 & 6.1--15\\
$f_{\mathrm{unabs}}$ $\left(10^{-14}\,\mathrm{erg\, s^{-1}\, cm^{-2}}\right)$\tablenotemark{c} & 6.7--77 & 1.7--7.5 & 3.6--21 & 1.5--4.9\\
\enddata

\tablenotetext{a}{Uncertainty ranges indicate 90\% confidence intervals.}

\tablenotetext{b}{Absorbed flux in the 0.5--2.0\,keV band. Range indicates 90\%
confidence interval.}

\tablenotetext{c}{Unabsorbed flux in the 0.5--2.0\,keV band. Range indicates 90\%
confidence interval.}
\end{deluxetable}

\begin{deluxetable}{lcccccccc}
\tabletypesize{\footnotesize}
\tablewidth{0pt}
\tablecolumns{9}
\rotate
\tablecaption{High-magnetic-field Radio Pulsars\label{tab:HighB}}
\tablehead{\colhead{Name} & \colhead{$P\,\left(\mathrm{s}\right)$} & \colhead{$B\,\left(10^{13}\,\mathrm{G}\right)$} & \colhead{$\dot{E}\,\left(\mathrm{erg\, s^{-1}}\right)$} & \colhead{$\tau\,\left(\mathrm{kyr}\right)$} & \colhead{$d$\tablenotemark{a}$\,\left(\mathrm{kpc}\right)$} & \colhead{$L_{\mathrm{X}}$\tablenotemark{b}$\,\left(\mathrm{erg\, s^{-1}}\right)$} & \colhead{Energy Range (keV)} & \colhead{Reference}}

\startdata
J1847--0130 & 6.71 & 9.4 & $1.7\times10^{32}$ & 83 & 8.4 & $<5\times10^{33}$ & 2--10 & \citet{msk+03}\\
J1718--3718 & 3.38 & 7.4 & $1.6\times10^{33}$ & 34 & 4.5 & $0.2\textrm{--}6\times10^{33}$ & Bolometric & \citet{km05}\\
J1814--1744 & 3.98 & 5.5 & $4.7\times10^{32}$ & 85 & 10 & $<6.3\times10^{35}$ & 0.1--2.4 & \citet{pkc00}\\
 &  &  &  &  &  & $<4.3\times10^{33}$ & 2--10 & \\
J1734--3333 & 1.17 & 5.2 & $5.6\times10^{34}$ & 8.1 & 6.1 & $0.1\textrm{--}3.4\times10^{33}$ & 0.5--10 & This work\\
 &  &  &  &  &  & $0.03\textrm{--}2.2\times10^{32}$ & 2--10 & \\
J1819--1458\tablenotemark{c} & 4.26 & 5.0 & $2.9\times10^{32}$ & 117 & 3.6 & $2.8\textrm{--}4.3\times10^{33}$ & 0.3--5 & \citet{mrg+07}\\
J1846--0258 & 0.33 & 4.9 & $8.1\times10^{36}$ & 0.9 & 6.0\tablenotemark{d} & $2.5\textrm{--}2.8\times10^{34}$\tablenotemark{f} & 0.5--10 & \citet{nsgh08}\\
 &  &  &  &  &  & $1.2\textrm{--}1.7\times10^{35}$\tablenotemark{g} & 0.5--10 & \citet{nsgh08}\\
J1119--6127 & 0.41 & 4.1 & $2.3\times10^{36}$ & 1.7 & 8.4\tablenotemark{e} & $1.9\textrm{--}3.2\times10^{33}$ & 0.5--7 & \citet{sk08}\\
J0847--4316\tablenotemark{c} & 5.98 & 2.7 & $2.2\times10^{31}$ & 790 & 3.4 & $<1\times10^{32}$ & 0.3--8 & \citet{kec+09}\\
J1846--0257\tablenotemark{c} & 4.48 & 2.7 & $7.1\times10^{31}$ & 442 & 5.2 & $<3\times10^{32}$ & 0.3--8 & \citet{kec+09}\\
B0154+61 & 2.35 & 2.1 & $5.7\times10^{32}$ & 197 & 1.7 & $<8\times10^{31}$ & 0.3--10 & \citet{gklp04}\\
B1916+14 & 1.18 & 1.6 & $5.1\times10^{33}$ & 88 & 2.1 & $\sim3\times10^{31}$ & Bolometric & \citet{zkgl09}\\
\enddata

\tablenotetext{a}{Unless otherwise noted, all distances were estimated from the dispersion measure of the source.}

\tablenotetext{b}{The ranges for PSRs J1718--3718 and J1819--1458 indicate 68\% confidence intervals, while the ranges for PSRs J1734--3333, J1846--0258, and J1119--6127 indicate 90\% confidence intervals.}

\tablenotetext{c}{Pulsar classified as a rotating radio transient (RRAT).}

\tablenotetext{d}{The distance to PSR J1846--0258 was found from H~\textsc{i} and $^{13}\mathrm{CO}$ spectral measurements \citep{lt08}.}

\tablenotetext{e}{The distance to PSR J1119--6127 was found from H~\textsc{i} absorption measurements \citep{cmc04}.}

\tablenotetext{f}{This value is from 2000, prior to this pulsar's 2006 outburst.}

\tablenotetext{g}{This value is from 2006, during this pulsar's 2006 outburst.}
\end{deluxetable}


\begin{thebibliography}{}
\bibitem[Baring \& Harding(1998)]{bh98}Baring, M. G., \& Harding,
A. K\@. 1998, \apjl, 507, L55

\bibitem[Caswell et al.(2004)]{cmc04}Caswell, J. L., McClure-Griffiths,
N. M., \& Cheung, M. C. M\@. 2004, \mnras, 352, 1405

\bibitem[Cordes \& Lazio(2001)]{cl01}Cordes, J. M., \& Lazio, T.
J. W\@. 2001, \apj, 549, 997

\bibitem[Espinoza et al.(2011)]{elk+11}Espinoza, C. M., Lyne, A.
G., Kramer, M., Manchester, R. N., \& Kaspi, V. M\@. 2010, \nat,
submitted

\bibitem[Gavriil et al.(2008)]{ggg+08}Gavriil, F. P., Gonzalez, M.
E., Gotthelf, E. V., Kaspi, V. M., Livingstone, \& M. A., Woods, P. M\@.
2008, Science, 319, 1802

\bibitem[Georgakakis et al.(2004)]{ggv+04}Georgakakis, A., et al.
2004, \mnras, 349, 135

\bibitem[Gonzalez et al.(2004)]{gklp04}Gonzalez, M. E., Kaspi, V.
M., Lyne, A. G., \& Pivovaroff, M. J\@. 2004, \apjl, 610, L37

\bibitem[Gotthelf et al.(2004)]{ghbb04}Gotthelf, E. V., Halpern,
J. P., Buxton, M., \& Bailyn, C\@. 2004, \apj, 605, 368

\bibitem[Harding et al.(1999)]{hck99}Harding, A. K., Contopoulos,
I., \& Kazanas, D\@. 1999, \apjl, 525, L125

\bibitem[Jansen et al.(2001)]{jla+01}Jansen, F., et al. 2001, \aap,
365, L1

\bibitem[Kaplan et al.(2009)]{kec+09}Kaplan, D. L., Esposito, P.,
Chatterjee, S., Possenti, A., McLaughlin, M. A., Camilo, F., Chakrabarty,
D., \& Slane, P. O\@. 2009, \mnras, 400, 1445

\bibitem[Kaspi(2007)]{k07}Kaspi, V. M\@. 2007, \apss, 308, 1

\bibitem[Kaspi(2010)]{k10}Kaspi, V. M\@. 2010, PNAS, 107, 7147

\bibitem[Kaspi \& McLaughlin(2005)]{km05}Kaspi, V. M., \& McLaughlin,
M. A\@. 2005, \apjl, 618, L41

\bibitem[Kaspi et al.(2006)]{krh06}Kaspi, V. M., Roberts, M. S. E.,
\& Harding, A. K\@. 2006, in Compact Stellar X-ray Sources, ed.\ W.
H. G. Lewin \& M. van der Klis (Cambridge: Cambridge Univ. Press), 279

\bibitem[Kumar \& Safi-Harb(2008)]{ks08}Kumar, H. S., \& Safi-Harb,
S\@. 2008, \apjl, 678, L43

\bibitem[Leahy \& Tian(2008)]{lt08}Leahy, D. A., \& Tian, W. W\@.
2008, \aap, 480, L25

\bibitem[Livingstone et al.(2010)]{lkg10}Livingstone, M. A., Kaspi,
V. M., \& Gavriil, F. P\@. 2010, \apj, 710, 1710

\bibitem[McLaughlin et al.(2003)]{msk+03}McLaughlin, M. A., et al.
2003, \apjl, 591, L135

\bibitem[McLaughlin et al.(2007)]{mrg+07}McLaughlin, M. A., et al.
2007, \apj, 670, 1307

\bibitem[Mereghetti(2008)]{m08}Mereghetti, S\@. 2008, \aapr, 15,
225

\bibitem[Motch(2006)]{m06}Motch, C\@. 2006, in ESA Special Publication,
Vol. 604, The X-ray Universe 2005, ed.\ A. Wilson (Noordwijk: ESA
Publications Division), 383

\bibitem[Ng et al.(2008)]{nsgh08}Ng, C.-Y., Slane, P. O., Gaensler,
B. M., \& Hughes, J. P\@. 2008, \apj, 686, 508

\bibitem[Page et al.(2006)]{pgw06}Page, D., Geppert, U., \& Weber, F\@.
2006, \nphysa, 777, 497

\bibitem[Pivovaroff et al.(2000)]{pkc00}Pivovaroff, M. J., Kaspi,
V. M., \& Camilo, F\@. 2000, \apj, 535, 379

\bibitem[Pons et al.(2007)]{plmg07}Pons, J. A., Link, B., Miralles, J.
A., \& Geppert, U\@. 2007, \prl, 98, 071101

\bibitem[Safi-Harb \& Kumar(2008)]{sk08}Safi-Harb, S., \& Kumar,
H. S\@. 2008, \apj, 648, 532

\bibitem[Str\"uder et al.(2001)]{sbd+01}Str\"uder, L., et al. 2001, \aap,
365, L18

\bibitem[Spitkovsky(2006)]{s06}Spitkovsky, A\@. 2006, \apjl, 648,
L51

\bibitem[Thompson \& Duncan(1995)]{td95}Thompson, C., \& Duncan,
R. C\@. 1995, \mnras, 275, 255

\bibitem[Thompson \& Duncan(1996)]{td96}Thompson, C., \& Duncan,
R. C\@. 1996, \apj, 473, 322

\bibitem[Thompson et al.(2002)]{tlk02}Thompson, C., Lyutikov, M.,
\& Kulkarni, S. R\@. 2002, \apj, 574, 332

\bibitem[Treves et al.(2000)]{ttzc00}Treves, A., Turolla, R., Zane,
S., \& Colpi, M\@. 2000, \pasp, 112, 297

\bibitem[Turner et al.(2001)]{taa+01}Turner, M. J. L., et al. 2001,
\aap, 365, L27

\bibitem[Woods \& Thompson(2006)]{wt06}Woods, P. M., \& Thompson,
C\@. 2006, in Compact Stellar X-ray Sources, ed.\ W. H. G. Lewin
\& M. van der Klis (Cambridge: Cambridge Univ. Press), 547

\bibitem[Zhu et al.(2009)]{zkgl09}Zhu, W., Kaspi, V. M., Gonzalez,
M. E., \& Lyne, A. G\@. 2009, \apj, 704, 1321

\end{thebibliography}
\end{document}